\documentclass[aps,prl,twocolumn,amsmath,amssymb,superscriptaddress,showpacs,showkeys,preprintnumbers,nofootinbib]{revtex4-1}

\usepackage{dcolumn}
\usepackage{graphicx,color}
\usepackage{multirow}
\usepackage{textcomp}
\usepackage{physics}
\usepackage{dsfont}
\usepackage{pdfpages}

\usepackage{etoolbox} 

\makeatletter
\patchcmd{\@outputpage@head}{\@ifx{\LS@rot\@undefined}{}{\LS@rot}}{}{}{}
\makeatother

\usepackage{relsize}

\usepackage{tikz}	
\usetikzlibrary
{
	arrows,
	calc,
	through,
	shapes.misc,	
	shapes.arrows,
	chains,
	matrix,
	intersections,
	positioning,	
	scopes,
	mindmap,
	shadings,
	shapes,
	backgrounds,
	decorations.text,
	decorations.markings,
	decorations.pathmorphing,	
}
\begin{document}
  \newcommand {\nc} {\newcommand}
  \nc {\beq} {\begin{eqnarray}}
  \nc {\eeq} {\nonumber \end{eqnarray}}
  \nc {\eeqn}[1] {\label {#1} \end{eqnarray}}
  \nc {\eol} {\nonumber \\}
  \nc {\eoln}[1] {\label {#1} \\}
  \nc {\ve} [1] {\mbox{\boldmath $#1$}}
  \nc {\ves} [1] {\mbox{\boldmath ${\scriptstyle #1}$}}
  \nc {\mrm} [1] {\mathrm{#1}}
  \nc {\half} {\mbox{$\frac{1}{2}$}}
  \nc {\thal} {\mbox{$\frac{3}{2}$}}
  \nc {\fial} {\mbox{$\frac{5}{2}$}}
  \nc {\la} {\mbox{$\langle$}}
  \nc {\ra} {\mbox{$\rangle$}}
  \nc {\etal} {\emph{et al.}}
  \nc {\eq} [1] {(\ref{#1})}
  \nc {\Eq} [1] {Eq.~(\ref{#1})}

  \nc {\Refc} [2] {Refs.~\cite[#1]{#2}}
  \nc {\Sec} [1] {Sec.~\ref{#1}}
  \nc {\chap} [1] {Chapter~\ref{#1}}
  \nc {\anx} [1] {Appendix~\ref{#1}}
  \nc {\tbl} [1] {Table~\ref{#1}}
  \nc {\Fig} [1] {Fig.~\ref{#1}}
  \nc {\ex} [1] {$^{#1}$}
  \nc {\Sch} {Schr\"odinger }
  \nc {\flim} [2] {\mathop{\longrightarrow}\limits_{{#1}\rightarrow{#2}}}
  \nc {\textdegr}{$^{\circ}$}
  \nc {\inred} [1]{\textcolor{red}{#1}}
  \nc {\inblue} [1]{\textcolor{blue}{#1}}
  \nc {\IR} [1]{\textcolor{red}{#1}}
  \nc {\IB} [1]{\textcolor{blue}{#1}}
  \nc{\pderiv}[2]{\cfrac{\partial #1}{\partial #2}}
  \nc{\deriv}[2]{\cfrac{d#1}{d#2}}
\title{New perspectives on spectroscopic factor quenching from reactions}
\author{C.~Hebborn}
\email{hebborn@frib.msu.edu}
\affiliation{Facility for Rare Isotope Beams, Michigan State University, East Lansing, Michigan 48824, USA}
\affiliation{Lawrence Livermore National Laboratory, P.O. Box 808, L-414, Livermore, California 94551, USA}
\author{F.~M.~Nunes}
\affiliation{Facility for Rare Isotope Beams, Michigan State University, East Lansing, Michigan 48824, USA}
\affiliation{Department of Physics and Astronomy, Michigan State University, East Lansing, Michigan 48824, USA}\author{A.~E.~Lovell}
\affiliation{Theoretical Division, Los Alamos National Laboratory, Los Alamos, New Mexico 87545, USA}
\preprint{LLNL-JRNL-845215, LA-UR-23-21722}

\date{\today}
\begin{abstract}
 The evolution of single-particle strengths as the neutron-to-proton asymmetry changes informs us of the importance of short- and long-range correlations in  nuclei and has therefore been extensively studied for the last two decades. Surprisingly, the  
 strong  asymmetry dependence of these strengths {and their   extreme values for highly-asymmetric nuclei} inferred from  knockout reaction measurements {on a  target nucleus} are not consistent with what is extracted from electron-induced, transfer, and quasi-free reaction data, constituting a  {two-decade} old puzzle. This work presents the first consistent analysis of one-nucleon transfer and one-nucleon knockout data, in which theoretical uncertainties associated with the nucleon-nucleus effective interactions {considered} in the reaction models are quantified using a Bayesian analysis.  Our results demonstrate that, taking into account these uncertainties, {the spectroscopic strengths of loosely-bound nucleons extracted from both probes agree with each other and, although there are still discrepancies for deeply-bound nucleons, the slope of} the asymmetry dependence of the single-particle strengths inferred from transfer and knockout reactions are consistent within $1\sigma$.  Both probes {are consistent with a small asymmetry dependence of these strengths}.
The uncertainties obtained in this work represent a lower bound and are already significantly larger than the original estimates.
\end{abstract}
\maketitle

\noindent 
{\it Introduction:}
Systematic studies of nuclei along isotopic chains have revealed unexpected trends that challenge our understanding of nuclear structure~\cite{TANIHATA2013215,Otsuka20,NunesPhysToday}. 
While energy spectra hold an important component of this complex many-body puzzle,  reaction studies {since the 50s have been extracting} information on the composition of the nuclear wavefunction itself, and in particular the distribution of strength across various nuclear orbitals. {This is expressed in terms of a spectroscopic factor (SF), proportional to the probability that the system will be found in a particular configuration. These SFs are reduced compared to the independent particle model (IPM), due to   long-range correlations (LRC), associated mainly with pairing and  deformation effects, and short-range correlations (SRC). The  evolution of this shell structure away from stability therefore provides  unique insights on  correlations  in nuclei and on the fundamental nuclear force~\cite{Otsuka20,AUMANN2021103847}.  Moreover, because SRCs influence the equation of state~\cite{PhysRevC.91.025803},  are connected with the quark momentum distributions in nucleons bound inside nuclei~\cite{PhysRevLett.106.052301}, and affect lepton-nucleus interactions~\cite{doi:10.1146/annurev-nucl-102010-130255}, an accurate understanding of  SRCs will impact astrophysics, particle physics and neutrino physics. }

{The importance of these correlations in nuclei is quantified by comparing SFs extracted from experimental nucleon-removal data and   theoretical   predictions~\cite{AUMANN2021103847}.} 
For {two decades}, nuclear physicists have grappled with the asymmetry dependence of the 
ratio $\cal R$ between the SF extracted from experiment and that predicted by the nuclear shell model.
The now famous asymmetry plot showing $\cal R$ as a function of the difference between neutron and proton separations energies ($\Delta S$) has caused {significant debate}~\cite{AUMANN2021103847}. The asymmetry dependence of $\cal R$  found in the analysis of {one-nucleon knockout reactions  on a $^9$Be or $^{12}$C targets--often referred simply as knockout reactions--}\cite{Gade08,Gade14,Gade21} is not consistent with that found using other probes, namely for electron-induced~\cite{KRAMER2001267}, quasi-free~\cite{GOMEZRAMOS2018511} and transfer reactions~\cite{Tetal09,LeePRC06,PhysRevLett.111.042502} (see the recent review Ref.~\cite{AUMANN2021103847} for a full status). For the last {two decades}, many studies have attempted to understand the source of this inconsistency.  {Our work adds to these studies, although it brings a novel perspective.}

{SFs} are model dependent \cite{PhysRevC.92.034313,PhysRevC.104.034311}; { their} extraction  from experimental data require both a reaction model and a structure model.
The analysis of knockout reactions makes use of the eikonal reaction theory as well as large-scale shell-model calculations. To understand the asymmetry dependence of $\cal R$ associated with  knockout observables, the validity of the shell-model SFs and the eikonal model have been thoroughly analyzed, e.g.  Refs.~\cite{PhysRevLett.107.032501,PhysRevLett.103.202502,PhysRevC.92.014306,PhysRevC.104.L061301,ATKINSON2019135027,PASCHALIS2020135110}  discuss the importance  of SRCs and LRCs for structure predictions and Refs.~\cite{PhysRevLett.108.252501,PhysRevC.83.011601,Getal21,LB21,PhysRevC.46.2638,PhysRevC.100.054607,PhysRevC.78.054602} address the validity of the eikonal approximation.  Similarly, many studies  tested the validity  of the theories used in the transfer analyses~\cite{NunesTr11,PhysRevC.95.064608,PhysRevC.84.034607,PhysRevC.85.054621}, including benchmarks of the reaction models. 

Given that SFs are not observables, {a degree of caution needs to be taken in interpreting
	the results}. When using different probes, it is essential to make the same assumptions so the conclusions are {both comparable and reliable}. Equally necessary is a good understanding of the theoretical uncertainties without which any disagreement between results is rendered meaningless. One  earlier study did attempt to quantify the uncertainties associated with the reaction theory used in the transfer \cite{NunesTr11} however those estimates were obtained without a rigorous statistical analysis. 

Although in the majority of cases, the $ {\cal R}(\Delta S)$ plots contain only statistical errors from the experimental data, we understand there are significant uncertainties attributed to the reaction models themselves. Most noteworthy are the  uncertainties associated with the phenomenological fits of the effective interactions used, the so-called \textit{optical potentials}~\cite{WhitePaperOpticalPotentials}. Bayesian analyses of elastic scattering has led to the understanding that uncertainties associated with the optical potentials are larger than previously estimated ~\cite{Lovell_2021,PhysRevC.97.064612,CRetal22,KOUQ22}. 
{This work offers the first consistent analysis of knockout and transfer reaction data  on three different Ar isotopes using  Bayesian statistics to quantify the theoretical uncertainties associated with the nucleon-nucleus optical potentials}. \\

\noindent
{\it Methodology:} {We first reanalyze transfer data for the $^{34,36,46}$Ar$(p,d)$$^{33,35,45}$Ar(g.s.) reactions at $33A$~MeV ~\cite{PhysRevLett.104.112701}, using the Adiabatic Wave Approximation (ADWA)~\cite{JT74},  and take as input the nucleon-$^{A}\rm Ar$ and nucleon-$^{A-1} \rm Ar$ interactions at the beam energy and at half the deuteron energy for the incoming and outgoing channels, respectively.}
The nucleon-Ar optical potential parameters are directly sampled\footnote{{We sampled the multivariate Gaussian posterior distributions generated   by the python script provided in the Supplemental Material of Ref. \cite{KDUQ}}} from the recent global parameterization KDUQ~\cite{KDUQ} from which we compute the credible intervals for the transfer angular distributions using the code {\sc nlat}~\cite{NLAT}.

\begin{table}
\centering	\begin{tabular}{ccccc}
		\hline \hline
		& $S_n$ [MeV]&$J^\pi $& $(A-1)$ $J^\pi $ & $nlj$\\
		$^{32}$Ar&21.60&	$0^+$&5/2$^+$&0$d$5/2
\\		
		$^{34}$Ar&17.07&0$^+$&	$1/2^+$&1$s$1/2
\\
		$^{36}$Ar&15.26	&			0$^+$	&3/2$^+$&$0d3/2$
\\
		{$^{46}$Ar}&8.07&			0$^+$&	$7/2 ^-$&$0f7/2$\\	\hline \hline
	\end{tabular}
	\caption{Properties of the single-particle wavefunction for  $^{32,34,36,46}$Ar:  the neutron separation energy ($S_n$), the spin and parity of the nucleus ($J^\pi$), of the $A-1$ core ($(A-1)$ $J^\pi$) and the number of nodes $n$, the partial wave $l$ and the spin $j$ of the core-neutron single-particle wavefunction.} \label{Tabspstructure}
\end{table}

{We reanalyze $^{32}$Ar$+^9$Be$\to$$^{31}$Ar(g.s.)$+X$ at $65.1A$~MeV~\cite{KO32Ar} and $^{34,46}$Ar$+^9$Be$\to$$^{33,45}$Ar(g.s.)$+X$ at $70A$~MeV~\cite{KO34Ar,KO46Ar}, using the eikonal method~\cite{G59,HM85,HT03,PhysRevC.78.054602}, and we quantify the uncertainties arising from the $n$-$^9$Be  target interaction only. }Because KDUQ is not appropriate for light targets, we follow the work done in Ref.~\cite{KOUQ22}, and instead we generate mock elastic angular distributions with  a realistic potential~\cite{Weppner18}. {We assign to these mock data an error of $10\%$, which is common for elastic-scattering experiments with stable beams.} Parameter posterior distributions are obtained from the Bayesian analysis of the $n$-$^9$Be target elastic scattering and are propagated to obtain credible intervals for the knockout momentum distributions. For the core-$^9$Be interaction, we use the optical limit with the parameters of Ref.~\cite{PhysRevC.77.034607}  and the density of $^9\rm Be$ approximated by two-parameter Fermi distributions~\cite{PhysRevC.66.014610}.  We do not include the uncertainties associated with the core-$^9$Be interaction as there are no elastic-scattering data on these systems or realistic potential to generate mock data and it has been shown that the uncertainties arising from this interaction are less significant~\cite{KOUQ22}. 
In both the transfer and knockout calculations, we do not include the spin-orbit force for convenience, since we expect it to have a negligible effect.

Critical to both  reaction calculations are the structure input: the exact same description is used for the single-particle structure of the isotopes involved. 
We use a Wood-Saxon potential with a radius {of  $R_R=r_0\,  (A-1)^{1/3}$ fm with $r_0=1.25$~fm }and  diffuseness of $a_R=0.65$~fm\footnote{{We have verified that our conclusions do not change when using a different geometry ($r_0=1.3$~fm and $r_0=1.2$~fm with $a_R=0.65$~fm) for this Woods-Saxon potential.}} and we fit its depth to the  neutron separation energy. Details concerning the relevant single-particle states  used in the transfer and knockout calculations are given in Table~\ref{Tabspstructure}.\\

\noindent

\begin{figure}[t]
\centering
\includegraphics[width=\linewidth]{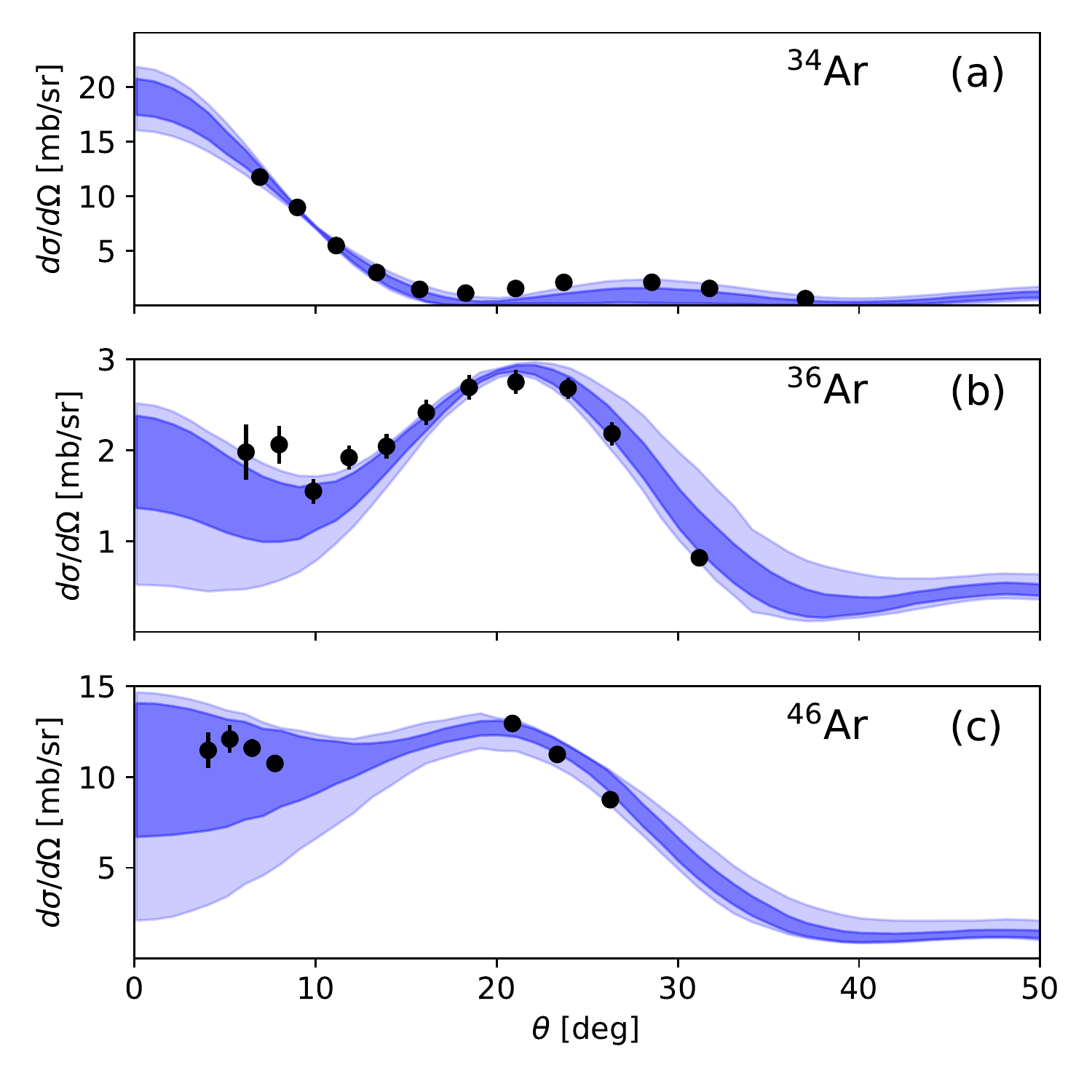}
\caption{Angular distributions for (a) $^{34}$Ar$(p,d)$$^{33}$Ar(g.s.), (b) $^{36}$Ar$(p,d)$$^{35}$Ar(g.s.) and (c) $^{46}$Ar$(p,d)$$^{45}$Ar(g.s.) at $33A$~MeV. All theoretical distributions have been scaled to reproduce the data from Ref.~\cite{PhysRevLett.104.112701}. The scaling factors are the extracted SFs {and their uncertainties result from both theoretical and experimental errors. }These values are listed in Table~\ref{Tab2}. } \label{Fig1}
\end{figure}

{\it Results:}
The transfer angular distributions are shown in Fig. \ref{Fig1}: the normalized 68\%  (dark shaded blue) and 95\% (light shaded blue) credible intervals  are compared to the data reported in Ref.~\cite{PhysRevLett.104.112701}. The shape of the predicted transfer angular distributions are in good agreement with experiment, corroborating  the assumptions made in ADWA.

\begin{figure}
    \centering
    \includegraphics[width=\linewidth]{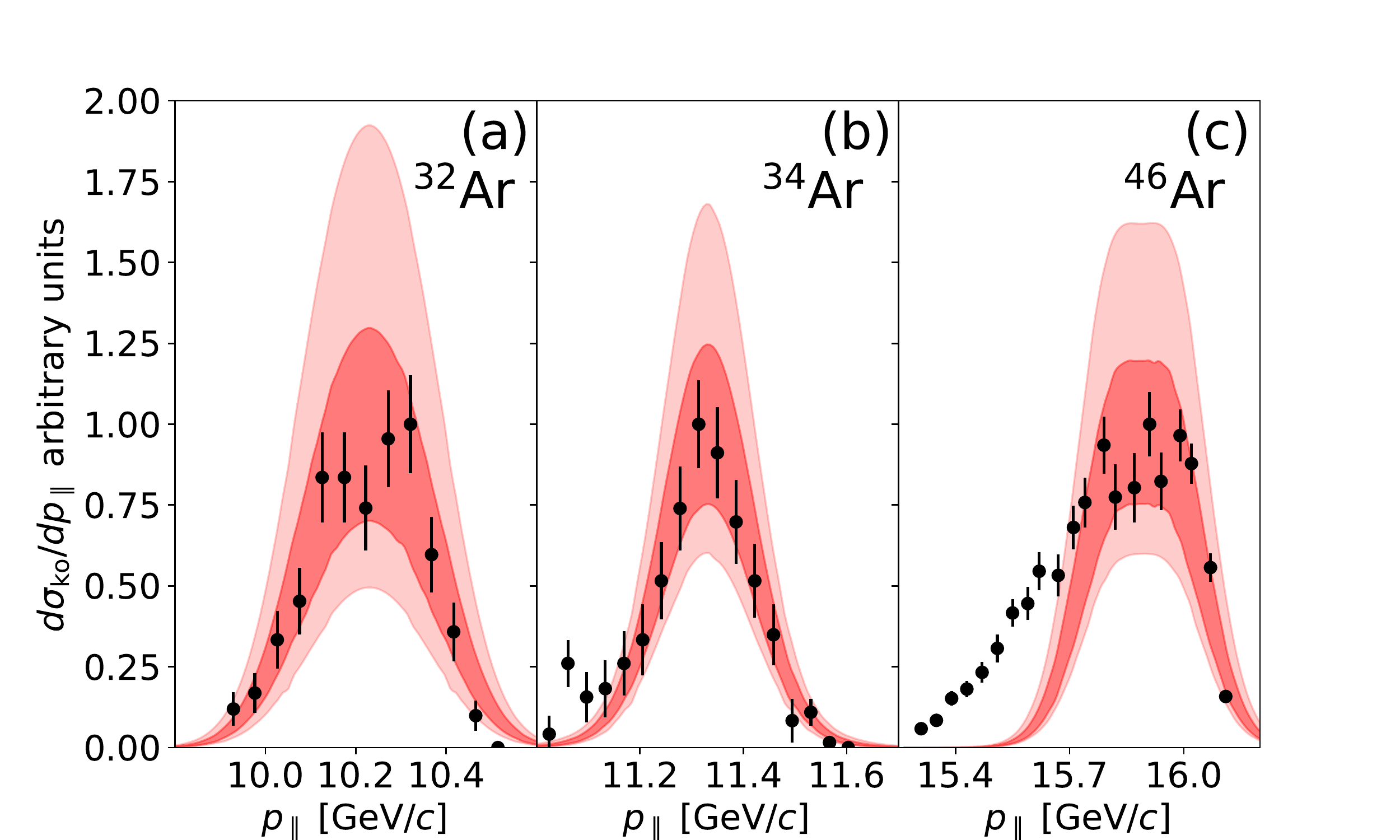}
    \caption{Parallel-momentum distributions of the remaining (a) $^{31}$Ar, (b) $^{33}$Ar and (c) $^{45}$Ar after the one-neutron knockout of $^{32}$Ar, $^{34}$Ar and $^{46}$Ar off a $^9\rm Be$ target at 65.1$A$~MeV, 70$A$~MeV and 70$A$~MeV, respectively. The theoretical distributions were
folded with the experimental resolution, and their center has been adjusted  so that the average distribution reproduces the high-momentum tail of the data.  The data and the experimental resolution profile were taken from Refs.~\cite{KO32Ar,KO34Ar,KO46Ar}.}
    \label{Fig2}
\end{figure}

The results for parallel-momentum distributions following knockout are shown in Fig.~\ref{Fig2}: the normalized 68\% (dark shaded salmon) and 95\%  (light shaded salmon) credible intervals are compared to the data reported in Refs.~\cite{KO32Ar,KO34Ar,KO46Ar}. In general, the experimental distributions are well reproduced by the eikonal model, except for the $^{46}$Ar case, which exhibits a highly-asymmetric distribution. As discussed in Ref.~\cite{KO46Ar}, this low-momentum tail is likely due to additional, dissipative  mechanisms acting in the final state of the reaction products, which are not included in the eikonal approximation. {We will discuss how this impacts the extracted SF later.}

The  parallel-momentum distributions are numerically integrated to obtain the total knockout  cross sections shown in Table~\ref{Tab1}: the experimental total cross sections ($\sigma_{exp}$, 2$^{\rm nd}$ column) are listed along with the theoretical ones, namely the diffractive-breakup contributions ($\sigma_{dif}$, 4$^{\rm th}$ column),  the stripping contributions\footnote{{The diffractive-breakup contribution corresponds to the reaction channel in which both the core and the neutron survive the collision, and the stripping to the channel in which the neutron is absorbed by the target. }} ($\sigma_{str}$, 5$^{\rm th}$ column) and the total predicted single-particle cross sections ($\sigma_{sp}=\sigma_{dif}+\sigma_{str}$, 3$^{\rm rd}$ column). It is clear that  the  errors on the  theoretical total  knockout cross section are also mostly defined by the uncertainties in the stripping component. This feature is expected for the removal of well-bound neutrons (all cases considered in this work have neutron separation energies $S_n>8$~MeV) which makes the reaction process more sensitive to the details of the optical potentials~\cite{KOUQ22}.

We now consider the extraction of the SFs. For transfer, we follow a similar procedure as in Ref.~\cite{NunesTr11}: we extract the SF by adjusting the angular distributions to the data points around the peak. The corresponding SFs and their uncertainties are displayed in the 2$^{\rm nd}$ column of Table~\ref{Tab2}, along with the $1\sigma$ ($2\sigma$) errors. Our  SF$_{tran}$ are consistent with  those extracted in Ref.~\cite{NunesTr11} (3$^{\rm rd}$  column) although they  exhibit larger errors.     The relative uncertainty in SF$_{tran}$ increases with the binding energy of the projectile, similarly to what was observed in knockout observables~\cite{KOUQ22}.
For knockout, we sample both the total cross section posterior distribution predicted by theory and the corresponding experimental total cross sections, assuming a normal distribution. We  then extract the distribution of the SFs  by  taking the ratio of the experimental samples with the theoretical ones.  The corresponding SF$_{ko}$, shown in the 4$^{\rm th}$  column of Table~\ref{Tab2}, are consistent with the ones extracted in the original analyses (5$^{\rm th}$  column)~\cite{KO32Ar,KO34Ar,KO46Ar}. However, the original uncertainties for SF$_{ko}$ are much smaller than those we obtained here, just as was found in the transfer case. Note that the SFs extracted from knockout data on $^{46}$Ar ($^{34}$Ar) are consistent with the ones extracted from the transfer data within $1\sigma$ ($2\sigma$). 
\begin{table}
    \centering
    \begin{tabular}{c| r | r r r |}
                  	&$\sigma_{exp}$ [mb] &$\sigma_{sp}$ [mb]	& $\sigma_{dif}$ [mb] 		&$\sigma_{str}$ [mb]			  \\ \hline
   $^{32}$Ar      &10.4$^{+1.3}_{-1.3}$ 	&  ${8.9^{+1.6(9.1)}_{-3.4(4.6)}}$&2.7$^{+0.4(2.0)}_{-1.0(1.1)}$	&6.3$^{+1.2(9.5)}_{-3.8(4.4)}$	\\
$^{34}$Ar      	&4.7$^{+0.9}_{-0.9}$		&12.8$^{+2.4(7.4)}_{-3.6(5.6)}$&4.2$^{+0.5(2.0)}_{-1.3(1.6)}$	&8.6$^{+1.6(7.6)}_{-4.5(5.6)}$ 	   \\   
   $^{46}$Ar       &61$^{+9}_{-9}$	&13.4$^{+1.8(7.9)}_{-4.2(6.2)}$	&4.2$^{+0.5(2.1)}_{-1.3(1.6)}$	&9.2$^{+1.7(8.3)}_{-4.4(6.1)}$	\\
    \end{tabular}
    \caption{The knockout experimental ($\sigma_{exp}$) and theoretical single-particle cross section ($\sigma_{sp}$) along with their diffractive-breakup ($\sigma_{dif}$) and stripping contributions ($\sigma_{str}$). The numbers are organized as $X_{-Z(Z')}^{+Y(Y')}$ where $X$ denotes the average value, $Y$ and $Z$ ($Y'$ and $Z'$) correspond respectively to the $1\sigma$ (2$\sigma$) uncertainties obtained by propagating the uncertainties due to the neutron-$^{9}$Be target interaction. }
    \label{Tab1}
\end{table}

To obtain the ratio $\cal R$, we use previously published large-scale shell model calculations\footnote{These   spectroscopic factors include the center-of-mass correction  $[A/(A-1)]^N$ with $N$ the major oscillator quantum number~\cite{PhysRevC.10.543,HT03}.}:  SF$_{SM}=4.39$ for $^{32}$Ar \cite{KO32Ar}, and SF$_{SM}=1.39, 2.22, 5.51$ for $^{34,36,46}$Ar \cite{PhysRevLett.104.112701}. No uncertainties have been estimated for these predictions. As noted in earlier analyses~\cite{Gade08,Gade14,Gade21}, the shell-model SFs are significantly larger than the   SFs extracted from the knockout of deeply-bound nuclei, e.g. $^{32,34}$Ar, but are in agreement with those extracted from transfer. {The fact that for some reactions, we have SF$_{tran}>$SF$_{SM}$ and SF$_{ko}>$SF$_{SM}$, highlights again that SFs are model construct. In  theoretical structure calculations,  SFs  are normalized to ensure the nucleon number conservation. However, when extracted from experimental data,   SFs do not ensure the nucleon number conservation as the theoretical model used to analyze the data inconsistently treats  structure and reaction properties.}

\begin{figure}[t]
    \centering
    \includegraphics[width=\linewidth]{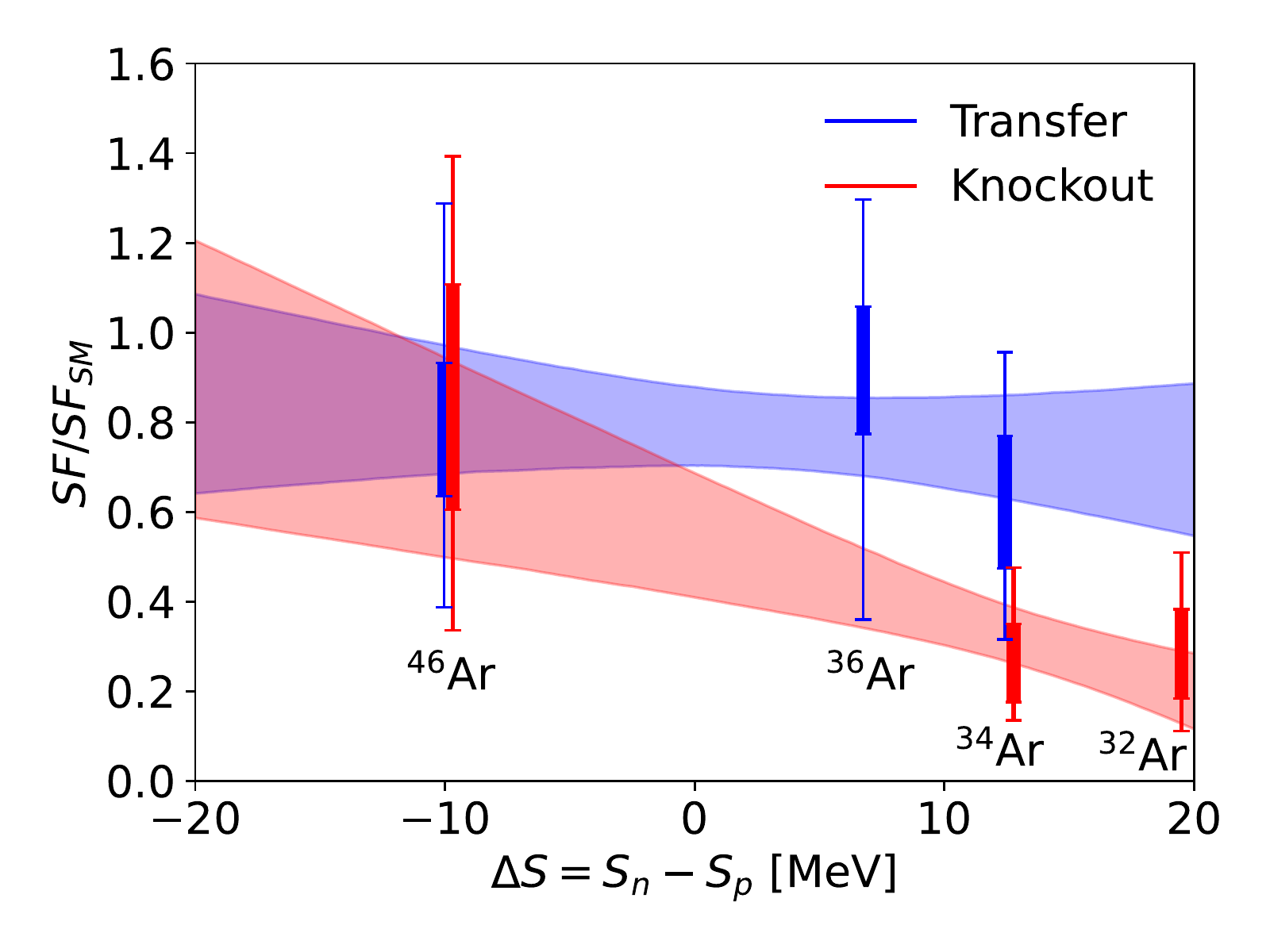}
    \caption{Ratio of the SF extracted from data  and the shell-model SF (including the center-of-mass correction) as a function of the asymmetry of the nucleus $\Delta S=S_n-S_p$. The blue error bars correspond to the SFs extracted from transfer data~\cite{PhysRevLett.104.112701} and the red ones to the SFs extracted from knockout data~\cite{KO32Ar,KO34Ar,KO46Ar}. Each error bars show the $1\sigma$ and $2\sigma$ uncertainties. The shaded area correspond to the $1\sigma$ uncertainties of a linear fit of the transfer (blue) and knockout (red) error bars.}
    \label{Fig3}
\end{figure}

Fig.~\ref{Fig3} contains the asymmetry dependence of  $\cal R$  using transfer (blue bars) and knockout (red bars) data. 
The thick bars represented $1\sigma$ and the thin bars represent $2\sigma$, both obtained from the credible intervals on the SFs reported in Table \ref{Tab2}. We must point out that the results for $\cal R$ obtained for the three discrete values of $\Delta S$ are not necessarily consistent with a linear dependence, either for transfer or for knockout. Nevertheless, for the sake of comparison with previous studies, we fit $\cal R$ to ${\cal R}(\Delta S)=a \Delta S + b$, for each reaction case, transfer or knockout, taking into account the $1\sigma$ uncertainty (blue and salmon shaded bands). 
The slope  obtained for transfer ($a=-0.0036 \pm 0.0090$) is consistent with the slope obtained for knockout ({$a=-0.0175 \pm 0.0091$)}, within uncertainties, even though there are significant differences in the values of the intercept (${\cal R}(0)=0.79 \pm 0.09$ for transfer and ${\cal R}(0)=0.55\pm 0.14$ for knockout). Our extracted slope for knockout is consistent with that extracted previously ($a=-0.016$~\cite{Gade21}) however now we include the $1\sigma$ uncertainty coming from both the optical potentials in the theoretical analysis and the  experimental errors. \\

\begin{table}
    \centering
    \begin{tabular}{c||cc|cc|c}
                  & SF$_{tran}$ & Ref. \cite{NunesTr11} & SF$_{ko}$ & Ref. \\ \hline 
   $^{32}$Ar      &&  &{1.3$^{+0.4(0.9)}_{-0.5(0.8)}$}&1.1$^{+0.1}_{-0.1}$ \cite{KO32Ar} \\
$^{34}$Ar      & 0.91$^{+0.16(0.42)}_{-0.25(0.47)}$& 0.92$^{+0.12}_{-0.12}$ &0.39$^{+0.08(0.23)}_{-0.16(0.25)}$&0.36$^{+0.07}_{-0.07}$ \cite{KO34Ar} \\   
   $^{36}$Ar      &2.1$^{+0.2(0.8)}_{-0.4(1.3)}$  &2.21$^{+0.49}_{-0.49}$ &&\\
   $^{46}$Ar      & 4.7$^{+0.5(2.4)}_{-1.2(2.5)}$ &4.93$^{+0.69}_{-0.69}$ &4.9$^{+1.3(2.8)}_{-1.5(2.8)}$&4.9$^{+0.7}_{-0.7}$ \cite{KO46Ar}\\
    \end{tabular}
    \caption{SFs extracted from transfer~\cite{PhysRevLett.104.112701} and knockout data~\cite{KO32Ar,KO34Ar,KO46Ar} compared with previous analyses ($3^{\rm rd}$ and $5^{\rm th}$ columns). The numbers are organized as $X_{-Z(Z')}^{+Y(Y')}$ where $X$ denotes the average value, $Y$ and $Z$ ($Y'$ and $Z'$) correspond respectively to the $1\sigma$ (2$\sigma$) uncertainties {obtained from the experimental errors and }by propagating the uncertainties due to the nucleon-nucleus interactions. }
    \label{Tab2}
\end{table}

\noindent
{\it Discussion:}
An optical model dependence on the overall normalization of the extracted SFs might be expected, but we have verified that the slope obtained is not dependent on the choice of the optical potential. When repeating the knockout analysis using KDUQ, the same parameterization used in the transfer analysis, we obtained  a slope of {$-0.0135 \pm 0.0124$} (shown in the Supplemental Material, which includes Ref.~\cite{KD03}). 
In addition to transfer and  knockout, $(p,pn)$ and $(p,2p)$ reactions have also been studied in this context~\cite{GOMEZRAMOS2018511}. Although to make a meaningful comparison, similar uncertainty analysis for those reaction channels needs to be done, our results do not seem inconsistent with Ref.~\cite{GOMEZRAMOS2018511}.

As pointed out in Ref.~\cite{GOMEZRAMOS2018511}, the asymmetry dependence extracted can change slightly when using shell-model predictions with different residual interactions and/or model spaces, or even when using different model assumptions for the geometry of the single-particle wave function. Yet, there will be no significant change on the relative  uncertainties due to the optical potentials. To remove any possible dependence on the shell model and to facilitate a future comparison with results obtained from $(e,e'p)$ measurements, we also provide the asymmetry plot  when  $\cal R$  is extracted using the IPM occupation numbers in Supplemental Material (the ratio  $\cal R$ deduced from the $(e,e'p)$ data~\cite{KRAMER2001267} relies on the IPM). The results for $\cal R$ using IPM do not seem inconsistent with those of Ref.~\cite{KRAMER2001267}, however a study of the uncertainty in that reaction probe remains to be completed. 

One must keep in mind that the uncertainties here presented are only a lower bound. First, we did not include the parametric uncertainties associated with the description of the single-particle state~\cite{CRetal22} even though we did use the exact same model in both analyses. For this reason, we expect these uncertainties to have no impact on our conclusions. Moreover, we did not quantify the uncertainties associated with the core-target potentials, used to compute knockout cross sections. More intricate is the quantification of model uncertainties. The reaction models used to interpret the measurements have approximations and therefore a complete analysis should include the errors associated with them. Although the inclusion of model uncertainties is beyond the scope of this work, there are plans to tackle this problem in the near future and it is important to identify the theory approximation in the models that are likely to be more relevant. Next, we briefly discuss this aspect.

In transfer reactions, ADWA has been benchmarked against  Faddeev calculations,  solving the three-body problem exactly~\cite{NunesTr11}: at $E_p=33$~MeV the differences are only significant for the $^{36}$Ar case but a more rigorous quantification is desirable. 
In knockout,  the measured momentum distributions exhibit an asymmetry for $^{46}$Ar that the model does not predict. The interpretation of the knockout data relies on the eikonal model, which contains two approximations: the  adiabatic approximation and a core-spectator approximation. The adiabatic approximation violates energy conservation and is the cause for the symmetric parallel-momentum distributions  (as seen in Fig.~\ref{Fig2}(c)). Improved models which do conserve energy are able to describe the distributions  (e.g., Ref.~\cite{PhysRevLett.108.252501}).
It has been shown that the integrated cross sections produced in such models agree with those from the eikonal model, proving that this aspect does not have a strong impact on the extracted SFs~\cite{Getal21}.   
The second approximation, the core-spectator approximation,  assumes that the core degrees of freedom are ``frozen'' during the collision process.  However, dissipative mechanisms associated with the removal of the nucleon tend to decrease the predicted cross sections~\cite{PhysRevC.83.011601}, an effect that is more important the more bound the system is.  We expect that the extracted SFs obtained from knockout data when including these dissipative effects would  be larger for  nuclei with large $\Delta S$, which could explain part of the apparent discrepancies between transfer and knockout predictions in Fig.~\ref{Fig3}. Unfortunately, accounting for these dissipative effects is not  trivial and requires updated reaction frameworks. Initial studies in the reaction theory community are moving in this direction~\cite{GOMEZRAMOS2022137252,GFK} but more work is needed, including the coupling of the new frameworks with a Bayesian analysis. \\

\noindent
{\it Conclusions:}
In summary, we reanalyze a set of transfer and knockout data using a Bayesian framework to quantify the theoretical uncertainties due to the optical potentials, known to be one of the leading sources of uncertainties in  reaction models. In the past, optical potentials uncertainties were estimated na\"ively by comparing the results with two arbitrary parameterizations~\cite{NunesTr11}. This work demonstrates that those original estimates produce uncertainties that are significantly underestimated. Most importantly, our results show that, when the optical potential uncertainties are included in a robust statistical approach, transfer and knockout reactions lead to a consistent picture {for the removal of a loosely-bound nucleon and both probes are consistent with a small    asymmetry dependence of SFs}. 
{This work also shows that there is still some tension between the strengths extracted from transfer and knockout data on deeply-bound nuclei as they only agree within $2\sigma$. These tensions come likely from model uncertainties that have not been quantified in this analysis, and will be the focus of future works.} Even though theoretical uncertainties need to be quantified in the analysis of $(p,2p)$, $(p,pn)$ and $(e,e'p)$ data to make a meaningful comparison between these probes and the present work, the slopes that we extract here do not seem inconsistent with previous analyses of those data~\cite{GOMEZRAMOS2018511,KRAMER2001267}.

Finally, it is also clear that to infer accurate and precise information from reaction data, optical potentials need to be better constrained. Of particular relevance {are} their imaginary strengths simulating the loss of flux from the elastic  channel due to open reaction channels {and their isospin dependence which drives the extrapolation of these interactions to unstable nuclei}.  To  improve  these phenomenological interactions, one can enforce the dispersion conditions,  relating   the real and imaginary parts of the  interaction in the appropriate manner~\cite{DICKHOFF2019252}, {and consider  other type of reaction data, e.g., charge-exchange, to constrain their isospin dependence}.  These dispersive potentials  provide a  consistent framework to describe the structure and reaction properties.  The extension of the Bayesian framework in this direction is being pursued.\\

\begin{acknowledgements}
\textit{Acknowledgments. } 
{C.~H. would like to thank L. Moschini for sharing her code computing the nuclear eikonal phase within the optical limit approximation and A. Gade for sharing the knockout data on $^{34}$Ar. Valuable discussions with T. R. Whitehead are acknowledged. C.~H. would like to thank K. Kravvaris, G. Potel  and  C. D. Pruitt for interesting discussions.} 
 C.~H. acknowledges the support of the U.S. Department of Energy, Office of Science, Office of Nuclear Physics, under the FRIB Theory Alliance award no. DE-SC0013617 and under Work Proposal no. SCW0498. A.~E.~L. acknowledges the support of the Laboratory Directed Research and Development program of Los Alamos National Laboratory. This work was performed under the auspices of the U.S. Department of Energy by Lawrence Livermore National Laboratory under Contract No. DE-AC52-07NA27344 and by Los Alamos National Laboratory under Contract 89233218CNA000001.
F. M. N. acknowledges the support of  the U.S. Department of Energy grant DE-SC0021422. This work relied on iCER and the High Performance Computing Center at Michigan State University for computational resources.
\end{acknowledgements}

\bibliographystyle{apsrev}
\bibliography{KOTransferUQ}
\onecolumngrid
\clearpage
 


\section*{Supplementary material}

# Supplemental Material for "New perspectives on spectroscopic factor quenching from reactions"

## A  Koning-Delaroche with uncertainty quantification (KDUQ)

The KDUQ [1] is the uncertainty-quantified version of the Koning-Delaroche (KD) global parametrization. It was fitted to the same corpus of data as KD, the main difference being the Bayesian framework used to determine the parameter posteriors and their correlations. It is valid for the scattering of a nucleon off a target with $A \geq 24$ and at energies $1$ keV $< E < 200$ MeV. In this section, we illustrate the posteriors and scattering observables obtained with the KDUQ.

For all calculations in this work, we neglect the spin-orbit terms in the KDUQ parametrization, that we expect to be small and to have a negligible effect of transfer and knockout observables. The KDUQ parametrization therefore includes only volume and surface terms, it reads

$$U(r) = -V f(r; R_V, a_V) - iW f(r; R_W, a_W) + i 4 a_S W_S \frac{d}{dr} f(r; R_S, a_S) \qquad (1)$$

where $f(r; R_i, a_i) = \frac{1}{1+e^{(r-R_i)/a_i}}$ and $V$, $W$, $W_S$ are the real volume, imaginary volume, and imaginary surface depths, $R_i = r_i A^{1/3}$ for the neutron-target potential, where $A$ is the mass number of the target.

Fig. 1(a) shows the corner plot obtained with 1600 sets of the KDUQ posteriors for $n$-$^{33}$Ar at 9 MeV. As noted in Ref. [1], the mean values of these posteriors are close to the parameters of the original KD (black line and points). By sampling these posteriors, one can propagate the uncertainties to reaction observables and build uncertainty bands on these cross sections. Fig. 1(b) shows the uncertainty bands obtained for the elastic-scattering cross section of a neutron of a $^{33}$Ar target at 9 MeV. Both the $1\sigma$ (dark blue shaded area) and $2\sigma$ (light blue shaded area) uncertainty bands include the prediction made using the KD (black line).

The uncertainty bands for the transfer observables [$^{34}$Ar$(p,d)^{33}$Ar(g.s.), $^{36}$Ar$(p,d)^{35}$Ar(g.s.) and $^{46}$Ar$(p,d)^{45}$Ar(g.s.) at 33$A$ MeV] shown in Fig. 1 of the manuscript are obtained by using 1600 sets of the KDUQ posteriors in the Adiabatic Distorted Wave Approximation (ADWA) [2]. The KDUQ is used for the $p$-$^{34,36,46}$Ar interactions at 33 MeV but also for the $n$- and $p$-$^{33,35,45}$Ar potentials at half the deuteron energies. Let us emphasize that each ADWA calculation uses three nucleon-nucleus potentials, that were built consistently from the same global parameter set (composed of 46 parameters [1]).

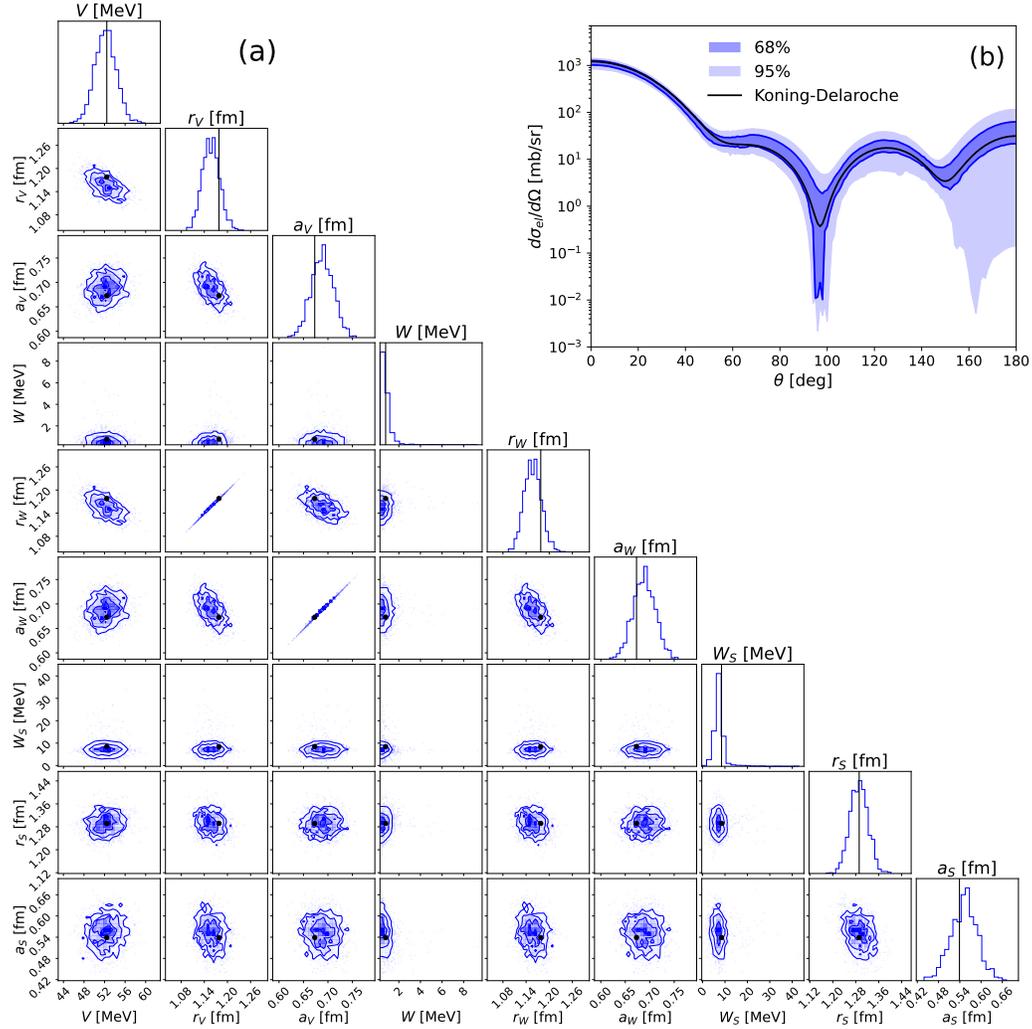

Figure 1: (a) Corner plot of the KDUQ [1] posteriors for $n$-$^{33}$Ar at 9 MeV ($V$, $r_V$, $a_V$, $W$, $r_W$, $a_W$, $W_S$, $r_S$ and $a_S$). (b) Elastic-scattering cross section of a neutron off $^{33}$ target. The dark and light blue shaded areas correspond respectively to the $1\sigma$ and $2\sigma$ uncertainty bands. The black line and points in (a) and (b) correspond to the original KD potential [3].

# B  Bayesian analysis of $n$-$^9$Be target interaction

In this work, we have propagated the uncertainties due to the neutron-target interaction onto knockout observables of $^{32,34,46}$Ar off a $^9$Be target. Since the KDUQ global optical potential is limited to heavier targets, its accuracy would be uncertain for $n$-$^9$Be system. To be able to quantify the uncertainties introduced in the fit of the potential parameters to experimental data for the $n$-$^9$Be interaction, we follow the methodology detailed in our previous work [4]. We consider mock data, that we generate at every 1° using a realistic potential [5], which was fitted to reproduce elastic-scattering observables for a nucleon off a target nucleus with $A \leq 13$ at energies between 65 MeV and 75 MeV. We assign an error of 10% to all mock data, which is common for elastic-scattering experiments with stable beams.

As in our previous work [4], we assume wide prior Gaussian distributions in order to obtain data-driven posterior, and we fix the geometry of the imaginary volume term to be the same as that of the real volume term, i.e., $r_W = r_V$ and $a_W = a_V$. The resulting parameters posteriors obtained for 25600 parameter sets, their correlations and the priors (black lines) used are shown in the corner plot Fig. 2(a). The $1\sigma$ (dark red shaded area) and $2\sigma$ (light red shaded area) uncertainty bands for the elastic-scattering observables along with the mock data are given in Fig. 2(b). The uncertainties on the scattering cross section are large at backwards angles, where no data were included in the fit. As seen in Ref. [4], including more data at these large angles would have a pronounced effect in the confidence intervals obtained for elastic scattering, but their impact on knockout observables is much more modest.

For each of the knockout calculations presented in the manuscript, we use 1600 samples of these posteriors to generate the uncertainty bands for knockout observables.

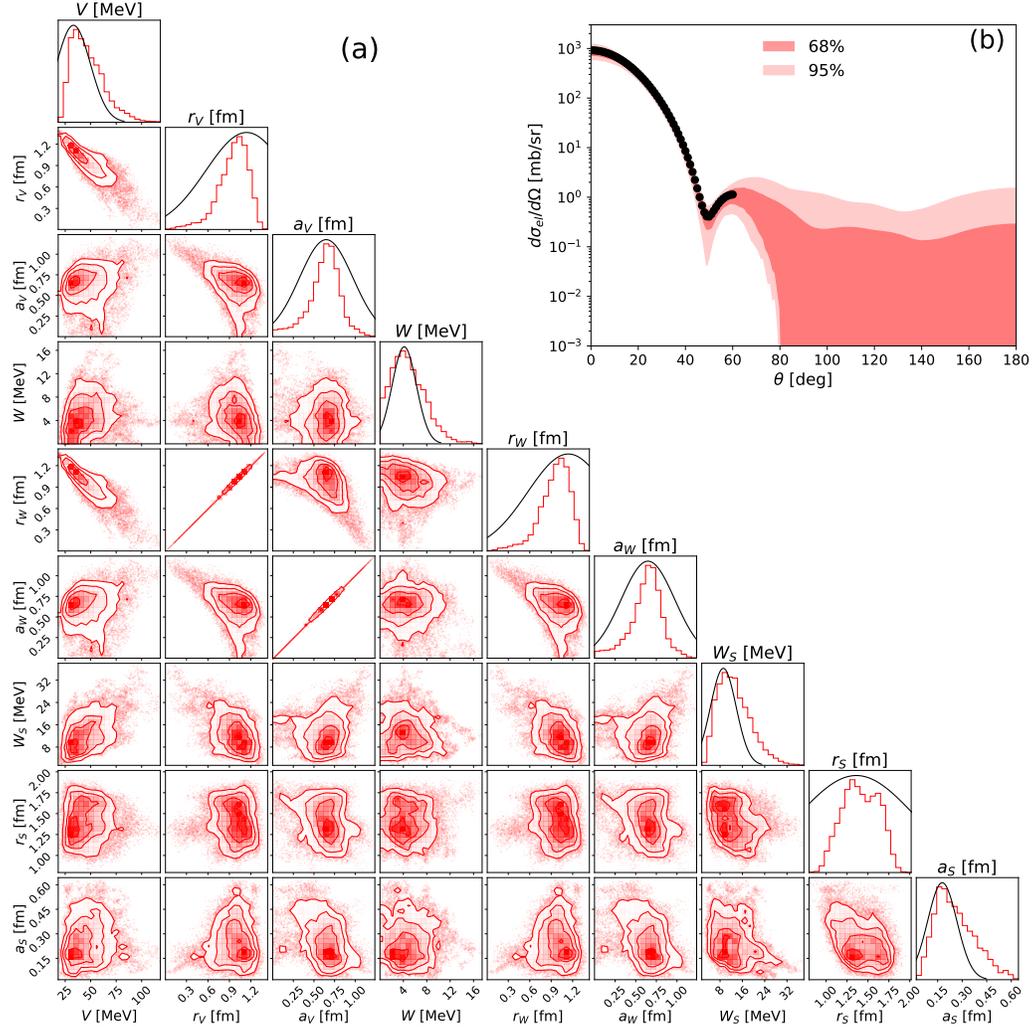

Figure 2: Phenomenological $n$-$^9$Be interaction constrained with mock data $\in [0, 60°]$ generated by the potential given in Ref. [5]. (a) Corner plots; (b) Elastic-scattering cross sections of $n$ off $^9$Be at 70 MeV as a function of the scattering angle $\theta$ with mock data $\in [0, 60°]$ (black points).

4

## C Asymmetry plot obtained with the KDUQ potential

In this section, we verify that the asymmetry dependence of the $\mathcal{R}(\Delta S)$ for knockout shown in Fig. 3 of the manuscript is not dependent on our methodology, i.e., the phenomenological Bayesian fit to $n$-$^9$Be mock data. We repeated our knockout calculations using the KDUQ potential for the $n$-$^9$Be interaction.

We show in Fig. 3 the resulting asymmetry plot of the ratio of the SF extracted from data and the shell-model SF. Similarly to what is seen in the manuscript, the SFs extracted from knockout and transfer observables are consistent within $2\sigma$. The slope of the knockout results ($a = -0.0135 \pm 0.0124$) is slightly smaller in magnitude and exhibits larger uncertainties than the one obtained in Fig. 3 of the manuscript ($a = -0.0175 \pm 0.0091$), exhibiting now a closer mean value to the one obtained from the transfer analysis ($a = -0.0036 \pm 0.0090$). These larger relative errors might be due to the fact that we extrapolate the KDUQ parametrization outside its mass range of validity. As the slope extracted from both knockout analyses are consistent, this indicates that the mild dependence of the ratio $\mathcal{R}(\Delta S)$ on the asymmetry of the nucleus is not dependent on how we construct the $n$-$^9$Be interaction.

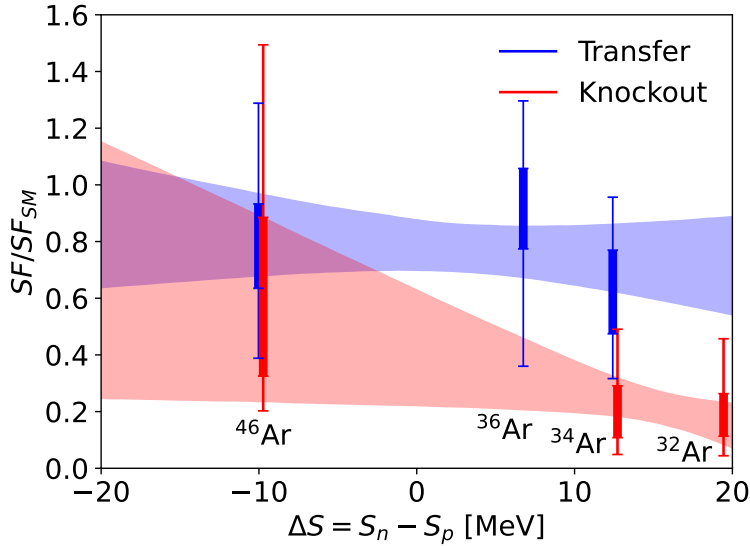

Figure 3: Ratio of the SF extracted from data and the shell-model SF (including the center-of-mass correction) as a function of the asymmetry of the nucleus $\Delta S = S_n - S_p$. The blue error bars correspond to the SFs extracted from transfer angular distributions and the red ones to the SFs extracted from knockout cross sections, both using the KDUQ parametrization [1]. Each error bars show the $1\sigma$ and $2\sigma$ uncertainties. The shaded area correspond to the $1\sigma$ uncertainties of a linear fit of the transfer (blue) and knockout (red) error bars. The slopes of these linear fits and their $1\sigma$ uncertainty are given in the text.

# D  Asymmetry plots obtained considering the independent-particle model occupation number

Because they change with the size of the model space and the interactions considered in the shell-model calculations, the slopes extracted from the asymmetry plot in Fig. 3 of the manuscript and Fig. 3 of this Supplemental Material are unique for each shell-model prediction. To avoid this ambiguity, one can extract the asymmetry dependence of the normalized ratio spectroscopic factors extracted from experimental data, obtained considering the independent-particle model (IPM) occupation number simply given by $(2j + 1)$[1]. Fig. 4 shows the asymmetry plots obtained from the analysis of experimental data using (a) the Bayesian analysis of the $n$-$^9$Be interaction for the knockout data and the KDUQ for the transfer ones, and (b) using in both cases the KDUQ parametrization.

Similarly to the case with the Bayesian analysis, the slopes obtained from both the knockout [(a) $a = -0.0118 \pm 0.0064$ and (b) $a = -0.0089 \pm 0.0087$] and transfer ($a = -0.0039 \pm 0.0061$) analyses are consistent, and the knockout calculations exhibit larger relative uncertainties. As discussed in the manuscript, the discrepancies between the knockout and transfer results are probably due to flaws of the eikonal model, that tends to overestimate the theoretical knockout cross sections and therefore underestimate the SFs extracted from knockout data.

Finally, even though a similar uncertainty analysis would need to be done to make a meaningful comparison with the $\mathcal{R}(0)$ deduced from $(e, e'p)$ data, our results do not seem inconsistent with the previous analysis [6]: we extract $\mathcal{R}(0) = 0.38 \pm 0.10$ from the knockout data in the case (a) [$\mathcal{R}(0) = 0.29 \pm 0.14$ in the case (b)] and $\mathcal{R}(0) = 0.51 \pm 0.06$ from the transfer data, while in Ref. [6] $\mathcal{R}(0) \sim 0.4 - 0.7$ is extracted from $(e, e'p)$ data.

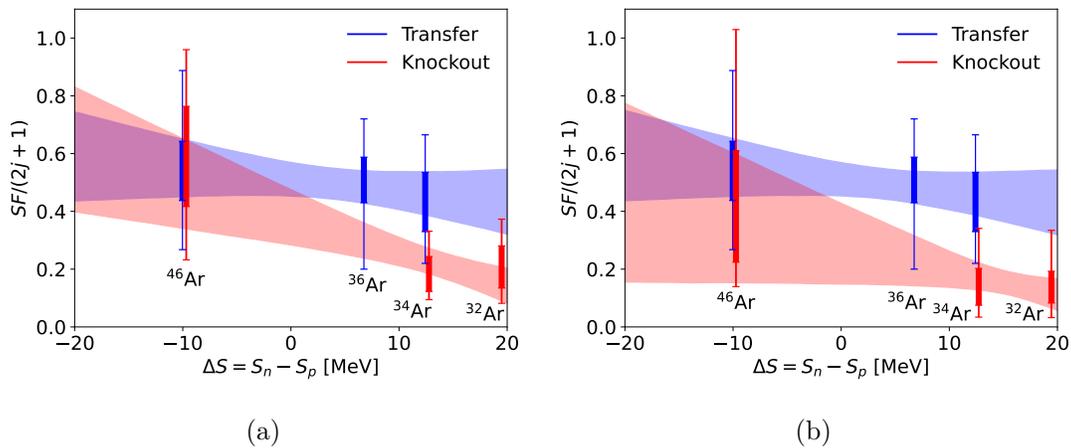

Figure 4: Asymmetry plot obtained with SFs extracted from transfer angular distributions (blue) and knockout cross sections (red) and normalized SFs considering the independent particle model. Panel (a) is obtained using the KDUQ parametrization for the transfer calculations and the Bayesian analysis of the $n$-$^9$Be interaction and panel (b) using in both cases the KDUQ parametrization. Each error bars show the $1\sigma$ and $2\sigma$ uncertainties. The shaded area correspond to the $1\sigma$ uncertainties of a linear fit of the transfer (blue) and knockout (red) error bars. The slopes of these linear fits and their $1\sigma$ uncertainty are given in the text.

---

[1] The IPM model by definition does not encapsulate any short- and long-range correlations (SRCs and LRCs). Since LRCs arise mainly from deformation and pairing, they are expected to be larger in open-shell nuclei, such as $^{36}$Ar, and the validity of the IPM is expected to be worse.



\end{document}